# MONITORING NORTH AFRICAN REGIONAL TOURISM BY WEB DATA

**Ilyes BOUMAHDI**

*Ph.D. at Gender, Economics, Actuarial, Statistics, Demography and Sustainable Development (GEAS3D) Laboratory, National Institute of Statistics and Applied Economics.*

iboumahdi@insea.ac.ma (Corresponding author)

**Nouzha ZAOUJAL**

*Professor-researcher at the National Institute of Statistics and Applied Economics, Rabat, Morocco.*

***Declarations:***

***1) Compliance with Ethical Standards:*** *This research study is in agreement with ethical standards as contained in the Committee on Publication Ethics guidelines.*

***2) Funding:*** *No funds, grants, or other support was received from any organization for the submitted manuscript.*

***3) Conflict of Interest:*** *On behalf of all authors, the corresponding author states that there is no conflict of interest.*

***4) Ethical Approval:*** *For this type of study, ethical approval is not required, as the data used didn't involve human or animal participants.*

***5) Informed Consent:*** *For this type of study, informed consent is not required.*

**Abstract**

The purpose of this article is to explore the opportunity of recent and detailed unconventional data from the tourism sector collected from « Booking.com » to make a finer and more up-to-date analysis than that established by conventional data, particularly, at the territorial level of North Africa.
We extracted and geolocalised about 40 variables of different types covering 1852 accommodations on Booking.com to analyze the characteristics of territorial tourist offer of the six North African countries (10 of 12 Moroccan regions, 3 of 13 Mauritanian Wilayas, 26 of 48 Algerian Wilayas, 13 of 24 Tunisian Governorates, 1 region of Libya, 15 of 27 Egyptian Mohafazats). Then, we used a random sample of 10% of the most recent appreciations of nearly 606000 tourists of the three most dynamic destinations (Marrakech-Safi, Tunis, Cairo) by analyzing the feelings of their comments with a differentiation according origin of tourists.
We concluded that the accommodation offer of the territories of North Africa is very diversified and unclassified offers are slightly better appreciated compared to those classified. The coastal regions have higher prices compared to the interior of the countries and quality-price appreciation of North African regions is below their overall ratings.

## 1. Introduction

The disruptive effect of information technologies has significantly reconfigured the way several sectors operate, including tourism (Buhalis, 2020; Fuchs & Sigala, 2020). This disruptive effect has gone from a standard digitalization phase to an acceleration phase since 2007 (Xiang, 2018) which has been accentuated following the massive adoption of smartphones (Mariani et al., 2019), immersive experiences caused by the Metaverse (Buhalis et al., 2023) and driven by the development of online platforms such as Airbnb (Bresciani et al., 2021; Chang & Sokol, 2022; Farmaki et al., 2020; Guttentag & Smith, 2017). However, several opportunities are emerging, including the exploitation of data provided by online tourist accommodation platforms, which are emerging as two-sided markets with major implications for the creation of value in the tourism sector (Dell'Era et al., 2021).

In addition to their commercial opportunities, these platforms would be an important tool for management (Park & Nicolau, 2017), forecasting (Wu et al., 2023) and monitoring and evaluation (Javornik et al., 2019; Liu et al., 2021) of the tourism sector through the data flows they generate. Indeed, the availability of multidimensional data capable of understanding and capturing the interactions underlying the tourism market is often lacking, particularly in developing countries and especially in times of crisis such as that of Covid-19 (Boumahdi et al., 2020). In addition, it would be difficult to better anticipate and prospect the evolution of the sector in order to support it with targeted public actions capable of supporting its development, in particular, at the territorial level. Indeed, the development and proliferation of information systems (online tourist accommodation platforms, reactions and assessments of tourists, data from professionals and support administrations, etc.), the vectors of their flows and the players in their exchanges make it possible to propose interesting alternatives to conventional data with as many challenges to overcome (very large flows, partial availability, complex analysis algorithms, representativeness, expertise, confidentiality, security, intelligence, ...) (Li et al., 2018; Lyu et al., 2022).

In this sense, we exploit the massive data of the website Booking.com in order to appreciate the territorial tourist offer of North Africa with the finest possible geographical coverage. Thus, this work provides an international perspective of the regional tourism offer of Morocco, in particular, and North Africa in general. We also highlight the most recent differentiated appreciation of consumers of the tourist offer of the territories of North Africa by circumventing the delay in publication and the limits of access to "conventional" data from surveys that are often very heavy to carry and very supervised. Thus, we collected and analyzed about forty variables of different types (numeric, alphanumeric, binary, geodesic, etc.) from the website Booking.com covering 68 territories of the six countries of North Africa (10 of the 12 regions of Morocco, 3 of the 13 Wilayas of Mauritania, 26 of the 48 Wilayas of Algeria, 13 of the 24 Governorates of Tunisia, 1 region of Libya, and 15 of the 27 Mohafazats of Egypt).

The ultimate objective of our study is to propose strategic axes best orienting local public actions in the tourism sector in North Africa. For this, we start by making a diagnosis of the tourist offer on the web of the territories of the six countries of North Africa and to raise the multidimensional appreciation made by the customers. In addition, we make a more detailed analysis of the three flagship destinations of North Africa, namely the Moroccan region of Marrakech-Safi, the Tunisian governorate of Tunis and the Egyptian Mohafazat of Cairo by analyzing the feelings and appreciation of the tourist clientele in relation to a sample accommodation.

## 2. *Method:*

We started by collecting data from 1852 accommodations in six North African countries (Morocco, Mauritania, Algeria, Tunisia, Libya, and Egypt) from the Booking.com tourist accommodation web platform using the extension of Google chrome "Data Miner"[1]. The data concerns around forty variables of different types (numeric, alphanumeric, binary, geodesic, etc.) for each accommodation. The data relates to a reservation made on February 3, 2020 for a stay of 6 nights (from August 1 to 7, 2020) for two adults. In order to inform the regional influence of each establishment, we created a geographic information system of the six countries of North Africa by concatenating six shapefiles of the territories of the six countries on Savgis (Souris et al., 1984). Next, we used the geo-membership module of the software SavGIS to collect the regional location of each settlement.

This approach made it possible to inform authomatically more than 90% of the regional location of accommodation establishments. The rest, which are close to the borders between the regions (Example of accommodation on the Nile between the mohafazats of Giza and Cairo), were informed manually. More accurate shapefiles at less than 10 meters precision would have facilitated the complete automation of georeferencing. Thus, the geographical allocation of the 1852 accommodations was allocated according to the 68 territories of the six countries of North Africa (10 of the 12 regions of Morocco, 3 of the 13 Wilayas of Mauritania, 26 of the 48 Wilayas of Algeria, 13 of the 24 Governorates of Tunisia, 1 region of Libya, and 15 of the 27 Mohafazats of Egypt) (Graph 1).

**Graph 1: Territorial tourist offer visible on the web of the six North African destinations in 2020**

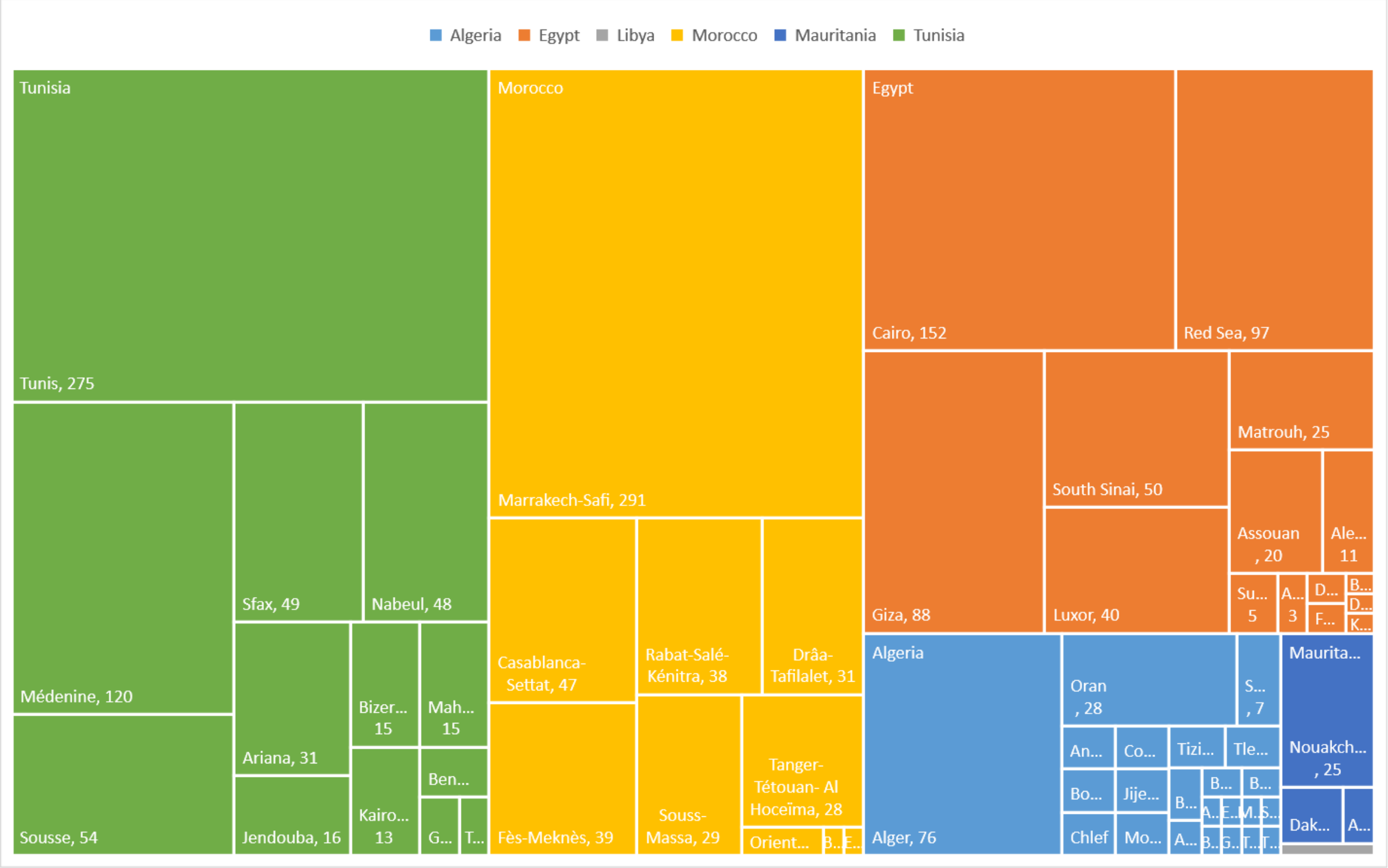

*Source: Authors from data extracted from the Web*

Then, we extracted the textual comments of tourists who actually stayed in the accommodations of the flagship destinations of North Africa, namely Marrakech-Safi, Tunis and Cairo, in order to make a more detailed analysis of the appreciation of these destinations. For this, we started by extracting, in a random way, 10% of the accommodations offered online from each territory, i.e. 72 accommodations (29 for the Marrakech-Safi region, 28 for the governorate of Tunis and 15 for the mohafazat of Cairo). Then, and for each accommodation, we extracted the ten most recent comments in order to have the most up-to-date appreciation of the tourist offer of these territories, i.e. more than 700 textual comments from tourists of 67 nationalities in about thirty languages, of which French is the most widespread. The analysis of the comments was made after correction of editorial errors, often carried out in non-formal spoken language, and translation into French[2].

[1] https://data-miner.io/

This peculiarity of writing comments on the web has been favorable to the textual analysis of feelings based on words and not on sentences. After cleaning up common words such as "a, absolutely, so,...", using the French version of stopwords-iso augmented with other words for the specific needs of this analysis, we did the textual analysis of the comments by R packages, such as tidytext, dplyr and tm (Feinerer, 2020). Sentiment analysis of comments depends on the software libraries used in the preparation of textual data and the routines used in their analysis, which often have been developed for English texts. For our study, we did the sentiment analysis using the French-speaking lexicon FEEL (Abdaoui et al., 2017). We also established the link of the most recurrent words with those having a positive or negative connotation by calculating the correlation between the two occurrence vectors of the two words according to their appearances in the 700 comments (Feinerer, 2020).

## *3. Results and discussion*

### *3.1. Characteristics of the territorial tourist offer on the web of North African destinations*

The territorial tourist offer, of North African destinations on the website Booking.com, is globally concentrated, exerting significant pressure on certain territories. The highest concentration concerns Moroccan destinations with 57% of accommodations in the Marrakech-Safi region, while it is the least important in Egyptian regions (31% in the Mohafazat of Cairo, 19% in that of the Red Sea and 18% in Giza) (Graph 2). The administrative division in Egypt has made it possible to relax the territorial concentration on the capital between the mohafazats of Cairo and Giza, at least on the administrative level. The concentration of accommodation is also higher in the Algerian Wilaya of Alger (48%) and in the Tunisian governorate of Tunis (42%). Compared to the area of the region, this concentration is the highest in the Tunisian governorate of Tunis with 85 accommodations per 100 square kilometer.

**Graph 2: Structure of the tourist offer visible on the web of the main North African tourist territories in 2020**

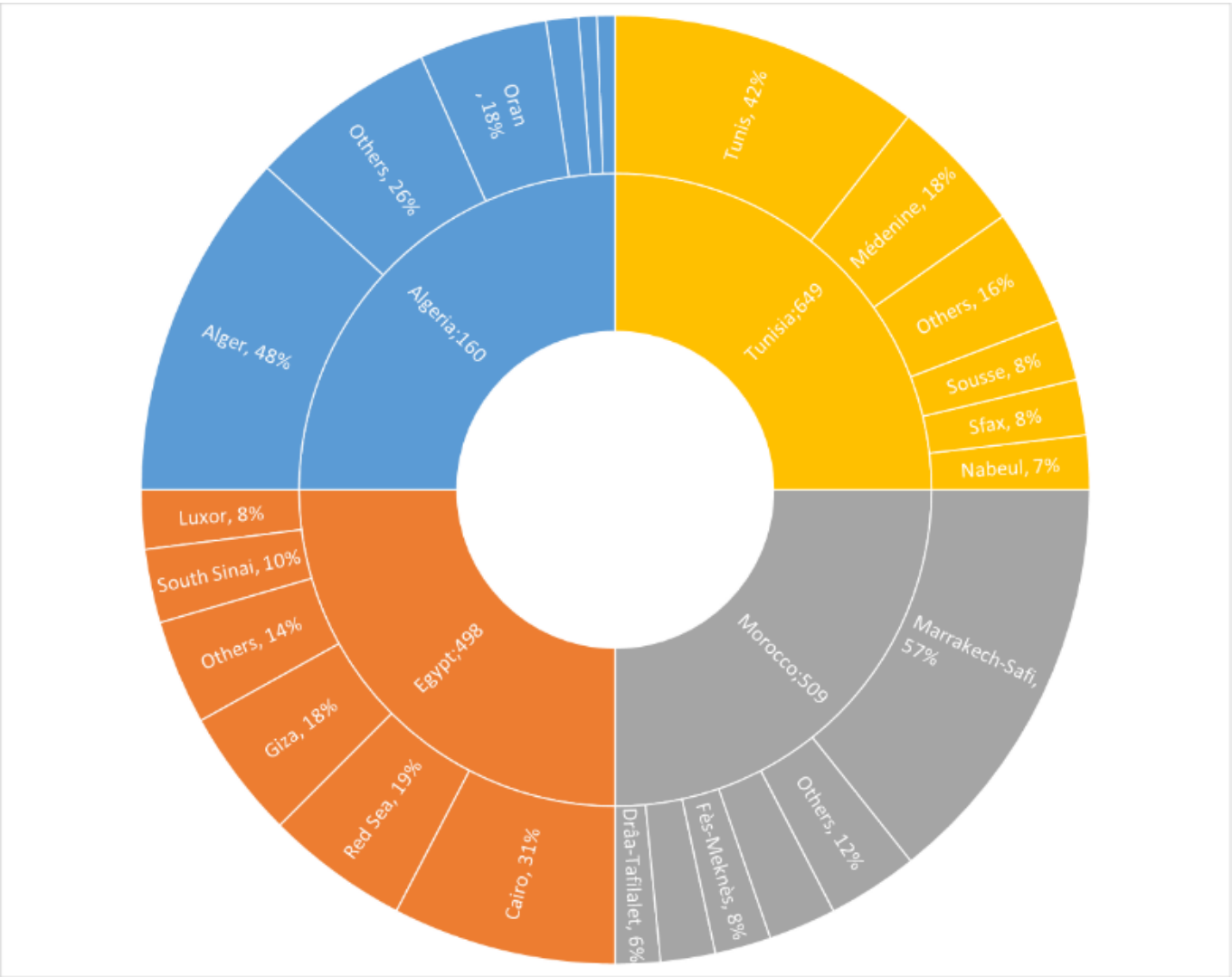

*Source: Authors from data extracted from the Web*

---

[2] *Manual processing of comments made it difficult to automate the translation. However, there are interesting solutions to explore in the future such as the translate Cloud API.*

The tourist offer on the web of the six North African destinations is characterized by the preponderance of hotel rooms (63%) while apartments make up 13% of supply on average (Graph 3). Supply for suite, studio and holiday home categories remains very low in North Africa with respectively 4%, 1% and 0.1% on average. Algeria has also a large offer in terms of apartments (28%), and especially in Alger (33%), while Morocco has a higher share of rooms (70%) than the North African average (63%). Territorially, a higher share of rooms in their accommodation offers (84%) characterizes the Moroccan region of Marrakech-Safi and the Egyptian Mohafazat of South Sinai while this share only reaches 55% in the Tunisian Governorate of Tunis.

**Graph 3: Territorial tourist offer visible on the web of the six North African destinations by type of accommodation in 2020**

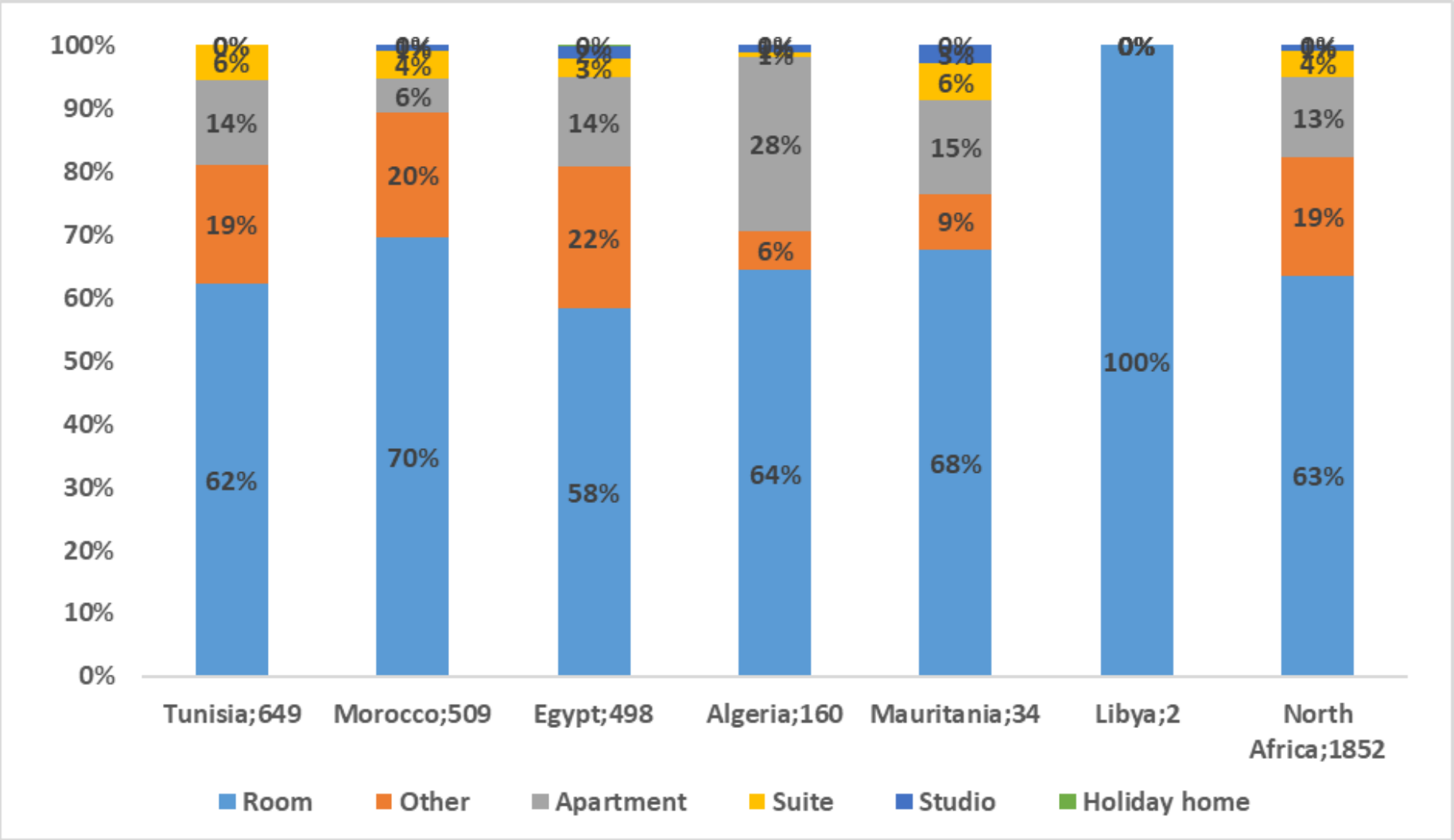

*Source: Authors from data extracted from the Web*

Nearly 69% of the tourist hotel offer in North Africa is unclassified with significant shares in the structure of that of Morocco (86%) and Egypt (80%) relative to Tunisia (54%) and Algeria (49%). The classified offer of North Africa is concentrated on the medium-high range up to 71% on average (respectively 37% in 5 stars and 34% in 4 stars). Tunisia has a slightly higher offer in the medium-high range (79%) while Algeria gives more importance to 3 stars (45%) (Graph 4). Territorially, the Moroccan region of Marrakech-Safi (43%), the Tunisian governorates of Souss (64%) and Sfax (78%), and the Egyptian Mohafazat of Cairo (57%) are characterized by a greater share of the 5 stars hotel class in their accommodation offers. On the other hand, the Algerian Wilaya of Alger has a classified offer concentrated on the category of 3 stars (51%).

**Graph 4: Classified tourist offer visible on the web of the six North African destinations by category in 2020**

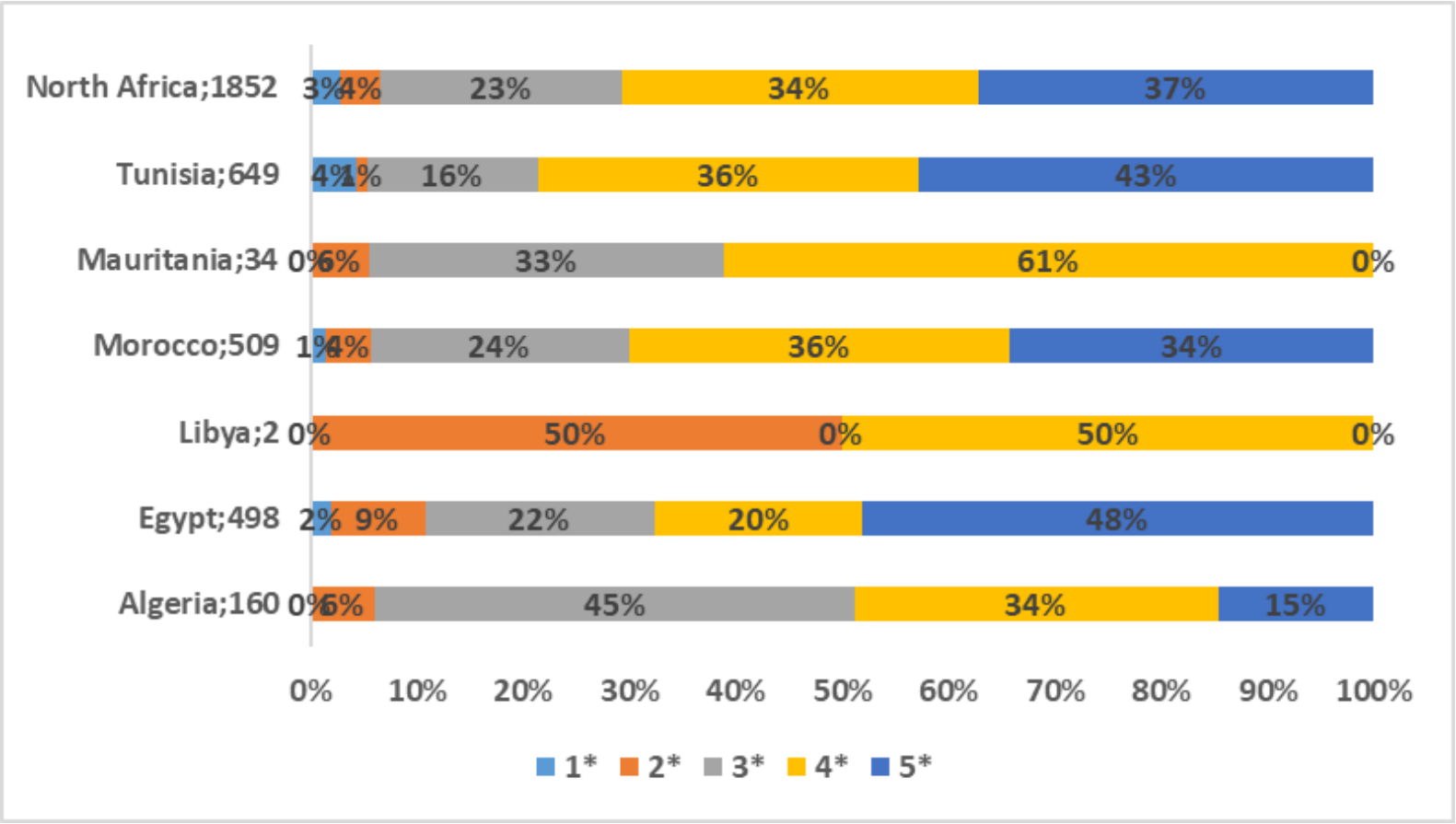

*Source: Authors from data extracted from the Web*

The conventional data observed by the official statistics of Morocco agree with the unconventional data that we collected from the web for the case of Marrakech-Safi. Indeed, the structure of the real capacity of Marrakech observed in 2018 is concentrated on the medium-high end up to 73% (respectively 43% in 5 stars and 33% in 4 stars) (Tourism observatory, 2019). Thus, we can confirm the integrity of our data, at least for the case of Marrakech-Safi, which is often controversial in the use case of big data. In addition, the average price of accommodation in all destinations in North Africa is €1406[3]. This average is pulled up by the high average tariff in Tunisia (€2770), particularly in the governorate of Nabeul (€3831). Prices in Morocco and Egypt are aligned at €675 with more dispersed prices in Egypt, the highest price being recorded at the Mohafazat of the Red Sea with €1054 against €505 for Giza. The lowest prices are recorded in Mauritania (€463) in the Wilaya of Adrar in particular (€144) (Graph 5).

**Graph 5: Average price of the territorial tourist offer visible on the web of the six North African destinations in 2020**

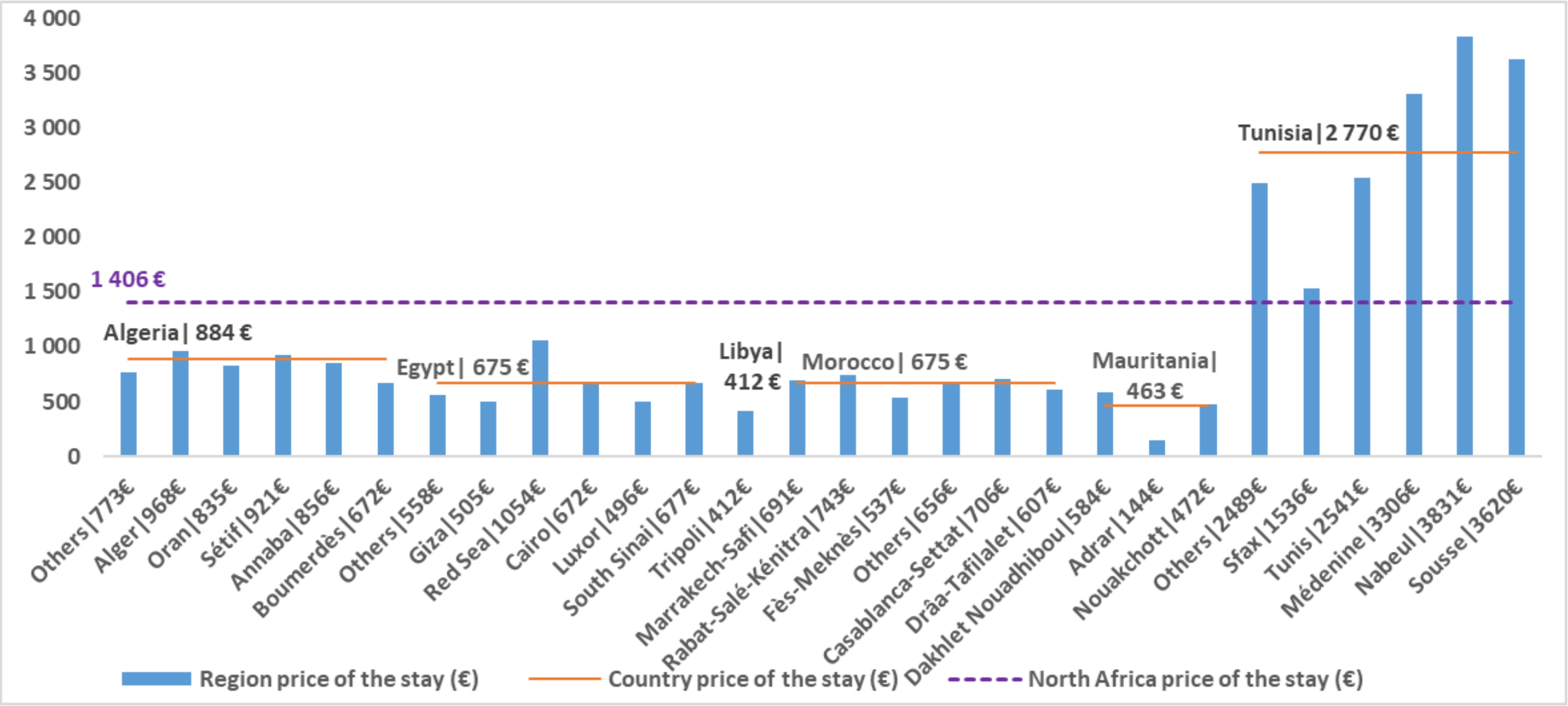

*Source: Authors from data extracted from the Web*

---

[3] *Average price for a booking of two adults made on February 3, 2020 for a six-night stay during the period from August 1 to 7, 2020.*

Thus, the coastal destinations seem to have higher prices compared to the interior of the countries (Graph 6). This is consistent with the finding that hotel room prices are about 10% higher for a room with a view of the Mediterranean compared to a room without a view specification (Fleischer, 2012). However, it should be noted that price differences could be due to the period of stay which coincides with the month of August which turns out to be a month of high season for certain coastal regions (Example of the Red Sea relative to Giza). The same is true for other price determinants such as holiday schedules, destination occupancy rates, and brand affiliation (Sánchez-Lozano et al., 2021). Finally, dark patterns tactics, such as not displaying additional charges, can be practiced, influencing consumers' cognitive biases as well as their decision-making (W. G. Kim et al., 2021).

**Graph 6: Map of the average price of the territorial tourist offer visible on the web of the six North African destinations in 2020**

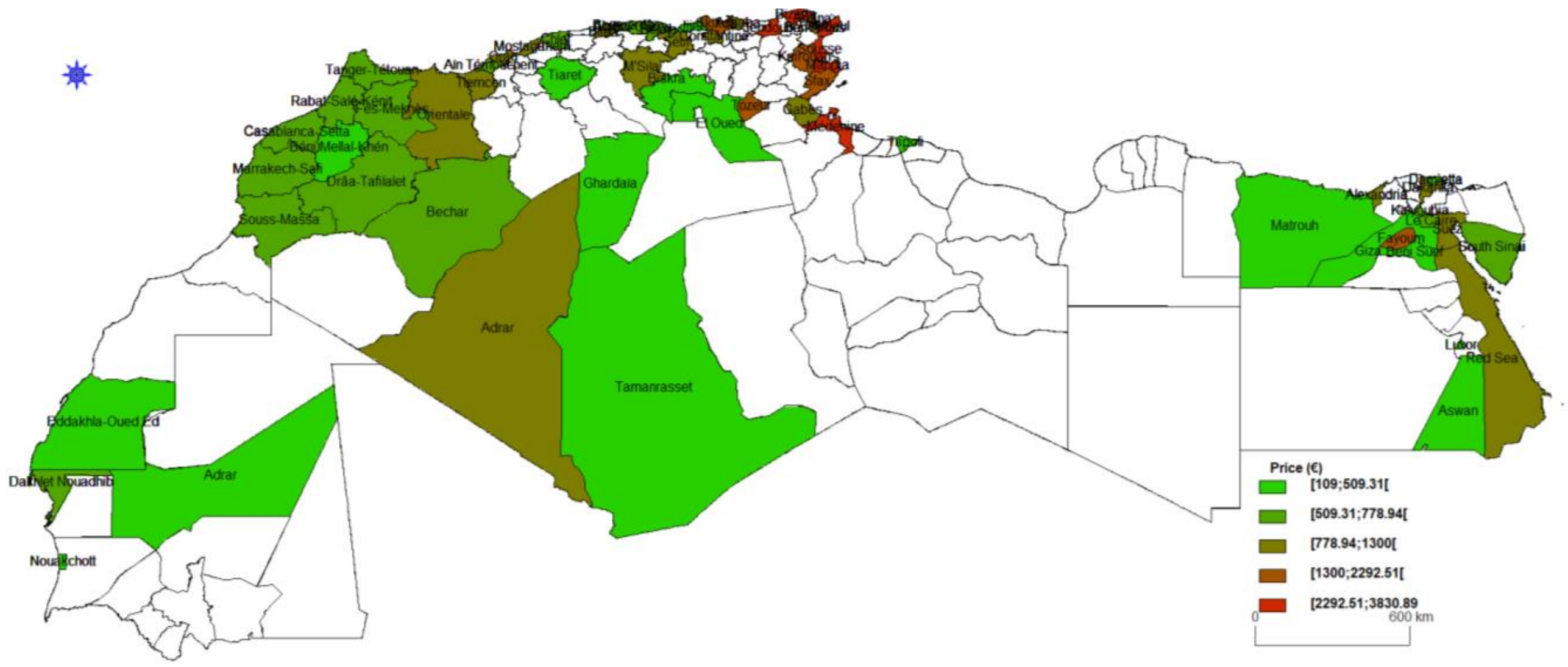

*Source: Authors from data extracted from the Web*

As for the tax burden, which depends on the legislature of local taxation and the structure of the hotel offer of each territory[4], it is on average[5] €46 in North Africa with an alignment of Moroccan destinations (€49) to Tunisians ones (€42) while it stands at €64 for Algeria (Graph 7). Cairo does not display specific taxes but all its accommodations reserve the possibility of adding additional charges. Maintaining tourist taxes at a low level makes it possible to keep prices at a competitive level with a beneficial attractiveness for the territory, but this to the detriment of its role in local finances, the financing of tourist promotion and the regulation of the tourist pressure on the territories (Cetin, 2014).

[4] *The tourist occupancy tax varies in the European Union between 0.1 € (Bulgaria) and 7.5 € (Belgium) per person and per night with an average range between 0.4 and 2.5 € (European Commission, 2023).*
[5] *Regarding a six-night stay for two adults.*

**Graph 7: Average tax of the territorial tourist offer visible on the web of the six North African destinations in 2020**

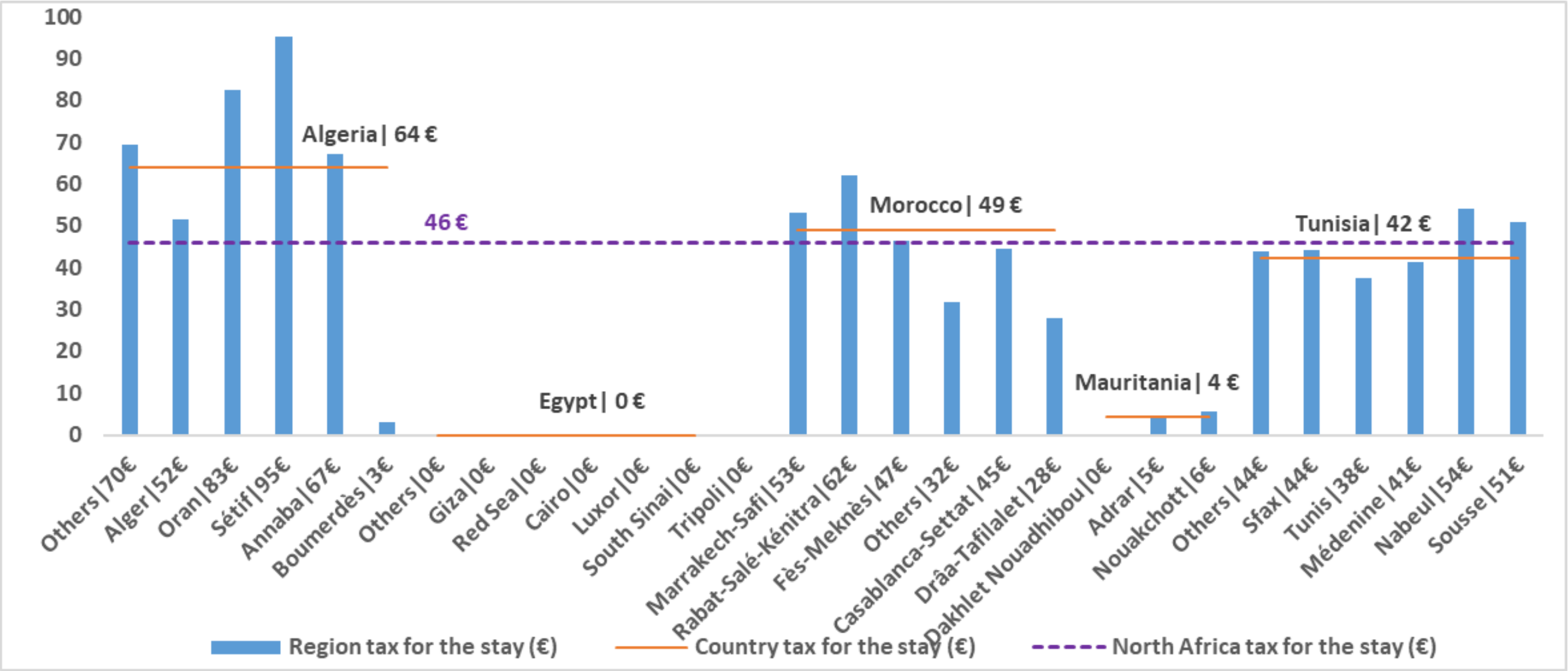

*Source: Authors from data extracted from the Web*

### *3.2. Appreciation of the territorial tourist offer on the web of the six North African destinations*

Based on the assessment on a 1 to 10 scale of nearly 606000 tourists who have had real experiences in terms of accommodation in the six North African destinations, it turns out that all the territories have a score similar to the average for North Africa (8.5) apart from Algeria (7.7) and Mauritania (7.6) (Graph 8). Unclassified offers from North Africa are slightly better appreciated (8.7) compared to those classified (8.0). These ratings relate to the year 2020, i.e. after the change in scoring made by the booking.com platform in September 2019 from a rating scale ranging from 2.5 to 10 to a rating scale ranging from 1 to 10. This will prevent us from having to resort to adjustment and harmonization techniques for future longitudinal analyzes (Amblee & Ullah, 2022).

**Graph 8: Assessment of the territorial web tourism offer of North African destinations (2020)**

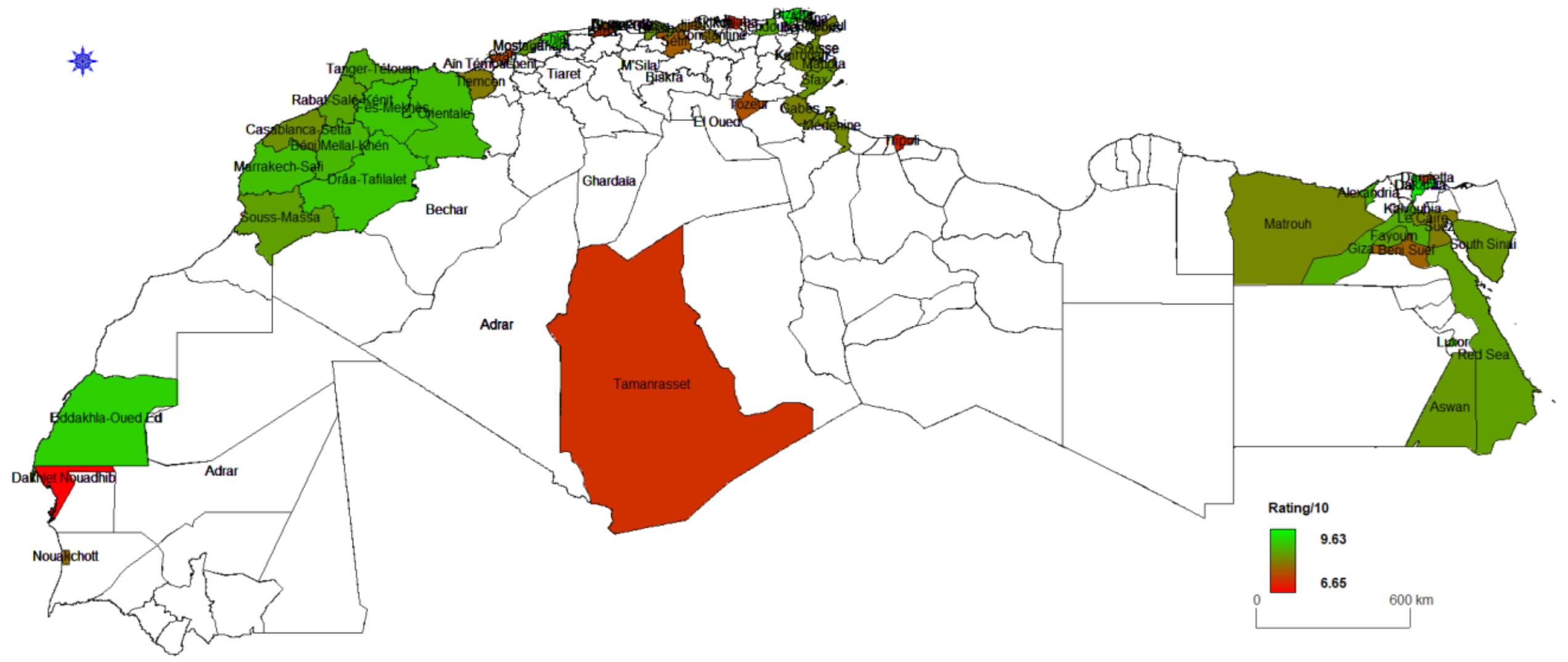

*Source: Authors from data extracted from the Web*

Morocco is slightly better appreciated (8.8), in particular, for accommodation in the Drâa-Tafilalet region, which is the best-rated destination in the territories of North Africa (8.9) (Graph 9). This general assessment is more homogeneous for all the destinations from the point of view of the geographical location of the establishments, since tourists gave them a similar average score (9.5). Territorial unclassified offers are slightly better appreciated compared to those classified, in particular, in Morocco at Drâa-Tafilalet (9.1), in Egypt at Giza (9.0) and in Tunisia in Sousse (9.0). Moreover, in the five-star category, Giza has an appreciation (7.8) below its average (8.7) and that of the same category for all North African destinations (8.2). This highlights the important role played by small reception structures in the attractiveness of territories and which should be supported in terms of funding and training. Indeed, if the increase in the size of the company would improve the functional and emotional experience of the customer, it would deteriorate his authentic experience (Ye et al., 2019) all the more if the commodification of the hosting service is important (Ye et al., 2018).

**Graph 9: Assessment of the territorial tourist offer visible on the web of the six North African destinations in 2020**

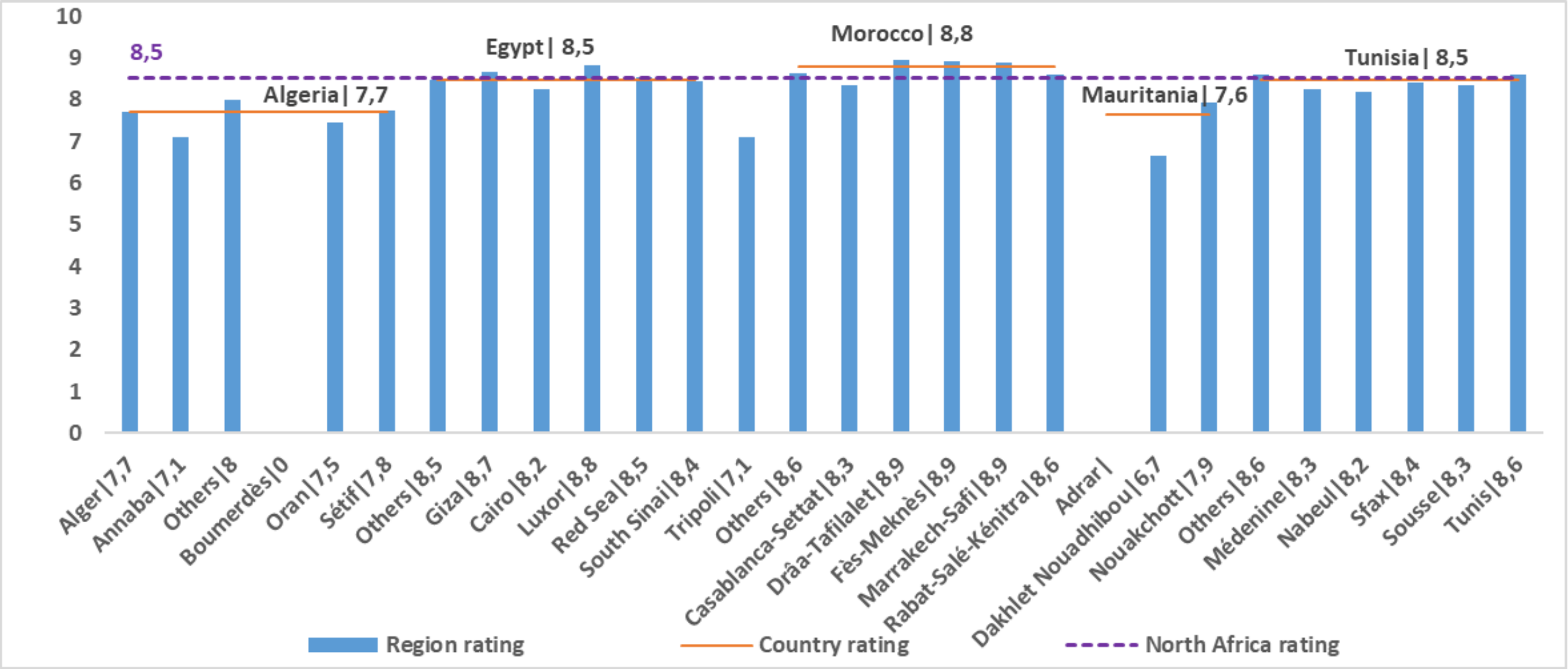

*Source: Authors from data extracted from the Web*

In order to decline the appreciation of the tourist territories of North Africa, we analyze the most important of them. These are those that offer more than 100 accommodations, namely Marrakech-Safi (291), Tunis (275) and Cairo (152). On the one hand, Cairo is better appreciated for the attention of its staff (8.6), cleanliness (8.3) and comfort (8.3) in the same way as the other flagship territories of the North Africa (respectively 9.4, 9.1 and 8.8 for Marrakech-Safi and 9.2, 9.0 and 9.0 for Tunis). However, it turns out that the overall assessment of the Mohafazat of Cairo (8.2), below that of Egypt (8.5) and North Africa (8.5), comes in particular to the appreciation of the quality/price ratio (7.8) in addition to the equipment (8.0), the breakfast (7.6) and the quality of the Wi-Fi (7.4) (Graph 10). However, Wi-Fi has become an asset whose absence, latency or limited access is a source of dissatisfaction (Mellinas & Nicolau, 2020). Moreover, on this point, the three territories have an appreciation that undermines their reputations with customers, while online reputation has gained in importance relative to the standard star rating (Abrate & Viglia, 2016).

**Graph 10: Thematic variation of the appreciation of the territorial tourist offer visible on the web of the flagship destinations of North Africa in 2020**

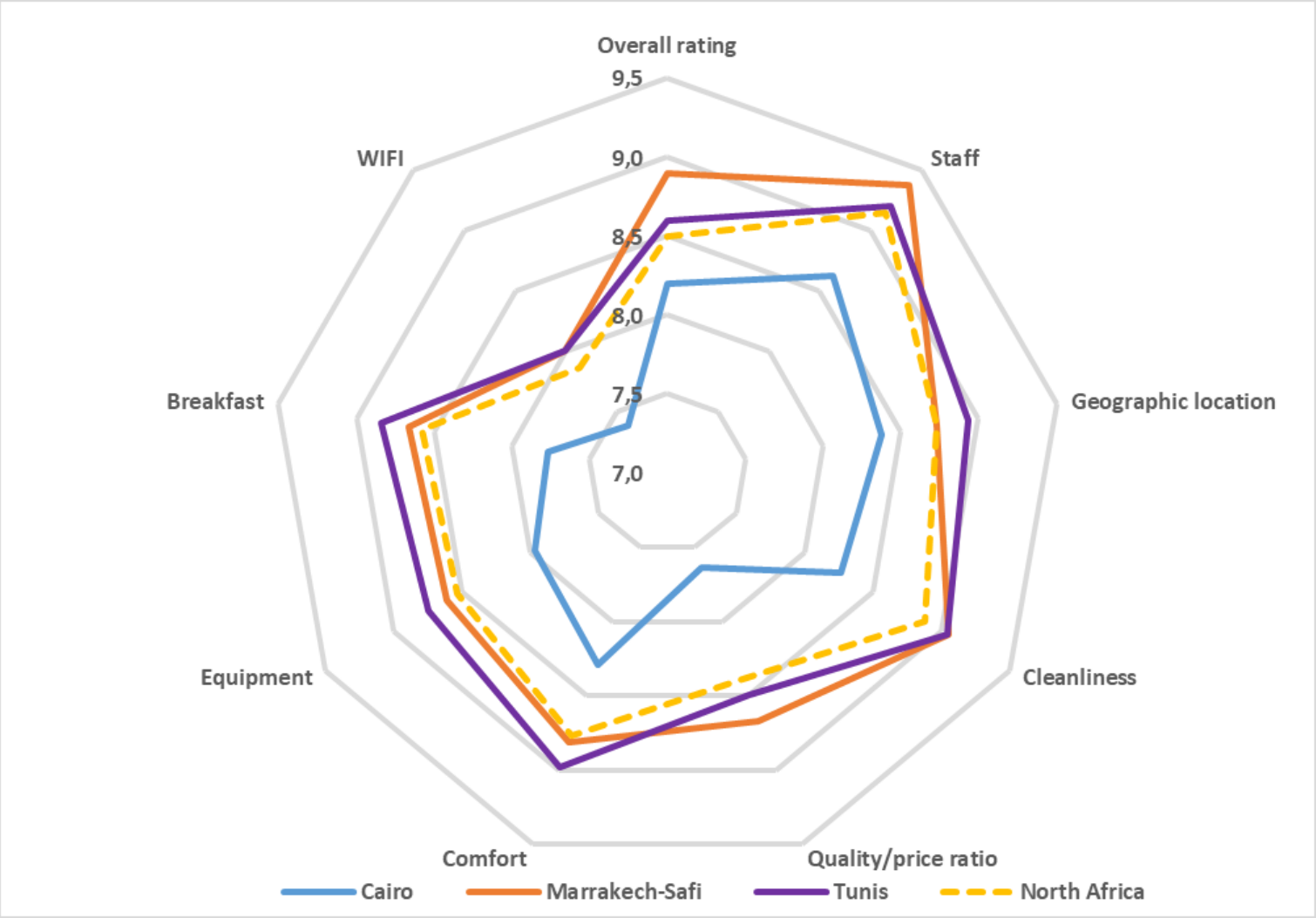

*Source: Authors from data extracted from the Web*

Thus, it turns out that for Cairo, in the same way as Tunis and Marrakech-Safi which also have an appreciation of the quality-price ratio below their overall ratings, tourists judge that the price paid is not appropriate their expectations in terms of quality of accommodation. Hat said, Cairo and Marrakech-Safi charge prices (675 € per stay) in the average of the country and relatively lower than those of Tunis (2541 €). Two alternatives would arise for these destinations, the first of which would be to reduce the price of the stay to create less expectation on the part of tourists and therefore more improved satisfaction given the price paid. Indeed, the non-confirmation of customer expectations has a negative effect on their accommodation ratings not only in terms of the evaluation of the quality-price ratio but also in that of the other characteristics of the accommodation (Abrate et al., 2021). The second alternative would be to improve the quality of service, in particular that linked to catering and the level of equipment, for Cairo and Marrakech-Safi and connectivity for the three destinations, in order to be up to tourist expectations.

National or regional support funds could be used to support professionals in this perspective with measures adapted to small structures and support in putting together their application file to benefit from these funds. Indeed, small hosting companies have difficulties in obtaining support from government institutions to develop and grow (Mxunyelwa & Henama, 2019). Moreover, some tourists are increasingly showing signs of relaxing their requirements with regard to certain aspects in favor of more ecological responsibility of hotel companies, particularly in terms of water and energy management. The green labeling of certain small hotel structures could relatively counter certain shortcomings, particularly in terms of comfort, cleanliness and equipment. Indeed, tourists would be willing to pay conventional hotel prices for a green hotel while tolerating minor inconveniences (reuse of towels, use of recycled products, etc.) (Y. Kim & Han, 2010). This is also verified for coastal destinations deploying the blue flag which improves their authenticity, destination brand identification, perceived quality and the tourist's predisposition to pay a higher price (Can et al., 2023). In addition, orientation towards the sustainability of accommodation improves customer satisfaction (Gerdt et al., 2019).

### *3.3. Detailed assessment of the tourist offer on the web of the flagship destinations of North Africa*

In order to make a more detailed analysis of the appreciation, we extracted from the web platform the textual comments of tourists who have actually stayed in the accommodation of the flagship destinations of North Africa, namely Marrakech-Safi, Tunis and Cairo. The interest of the analysis of comments is due to their influence on the attractiveness of territories since Internet users who were looking for reservations considered 9% of these comments useful. Moreover, the number of comments processed (700) seems to us sufficient to avoid the natural predisposition of people with dark personality traits to have negative online opinions (Yousaf & Kim, 2023). The structure of tourists is different between the three destinations with however the preponderance of France, Germany, Italy, Spain and the United Kingdom (Graph 11). Egypt stands out for the strong presence of nationals, which would compensate for the recent drop in inbound tourism activity. It is also distinguished by the strong presence of the Gulf countries, in particular, Saudi Arabia and the United Arab Emirates. Tunisia is distinguished by the presence of tourists from the border country (Libya) revealing the opportunity offered by the opening of borders and the free movement of tourists from the Maghreb.

**Graph 11: Structure of comments by country of origin of the territorial tourist offer visible on the web of the flagship destinations of North Africa in 2020**

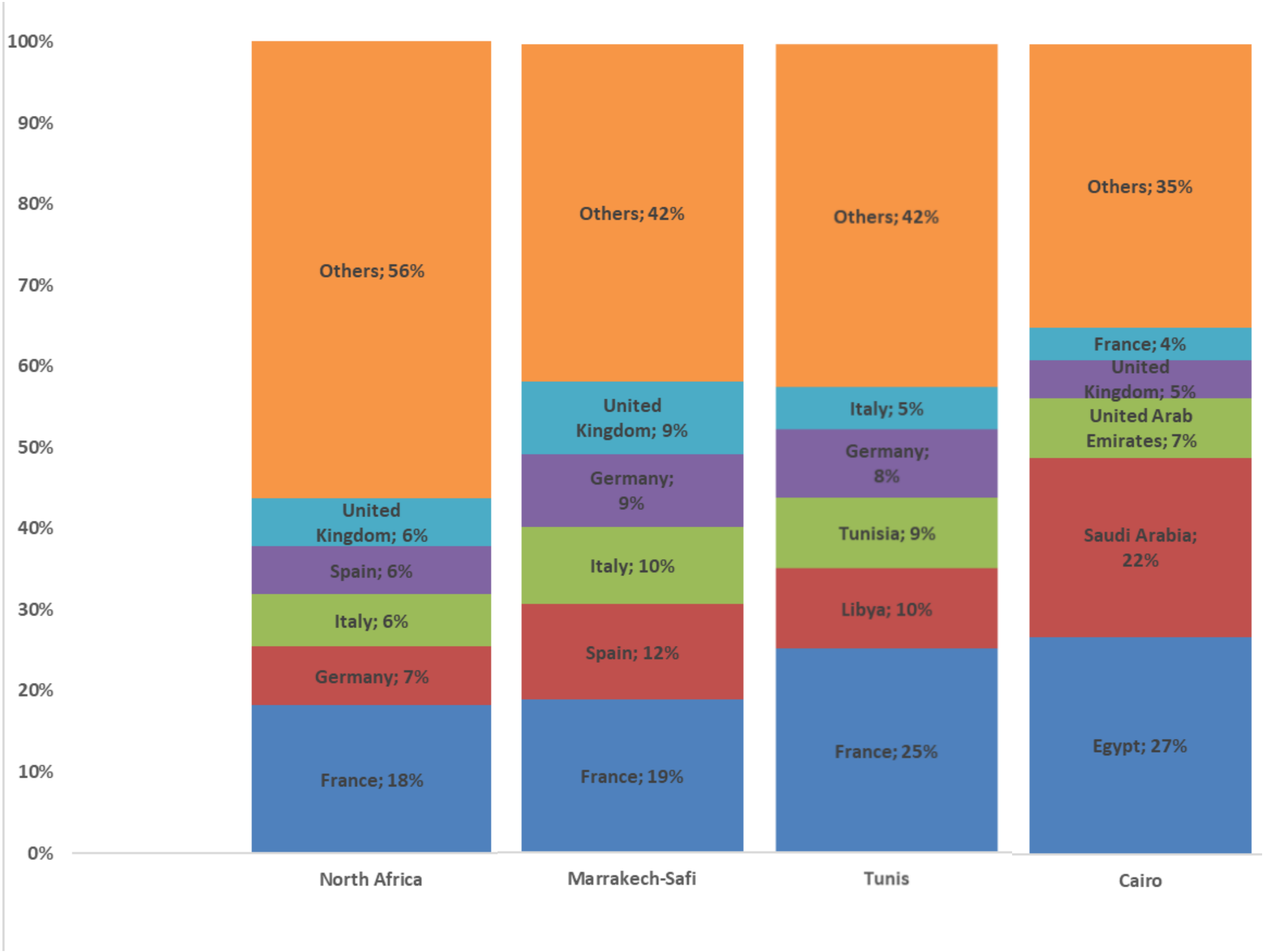

*Source: Authors from data extracted from the Web*

Moreover, these comments also seem to be representative if we compare their structures by nationality relative to the official regional statistics of Morocco for which we found data. Thus, the structure by origin of the tourist (58 nationalities) of the comments is similar in level to that noted by the conventional data in terms of arrivals and overnight stays for Marrakech-Safi, despite the difference in terms of use of the canals reservation by country[6]. Textual analysis of the comments reveals that the words that come up most often are “staff”, “breakfast”, “location” and “room” (Graph 12). This is consistent with other northern Mediterranean destinations such as Portugal where the terms that appear most frequently in reviews also relate to room, staff and location (Chaves et al., 2012).

---

[6] *France, United Kingdom, Germany and Spain are the four most important foreign source markets for Morocco and Marrakech-Safi concentrating respectively 46% and 40.1% of arrivals and respectively 56% and 49.5% nights in 2018 (Tourism observatory, 2019).*

**Graph 12: Word cloud of comments on the territorial tourist offer visible on the web of the flagship destinations of North Africa in 2020**

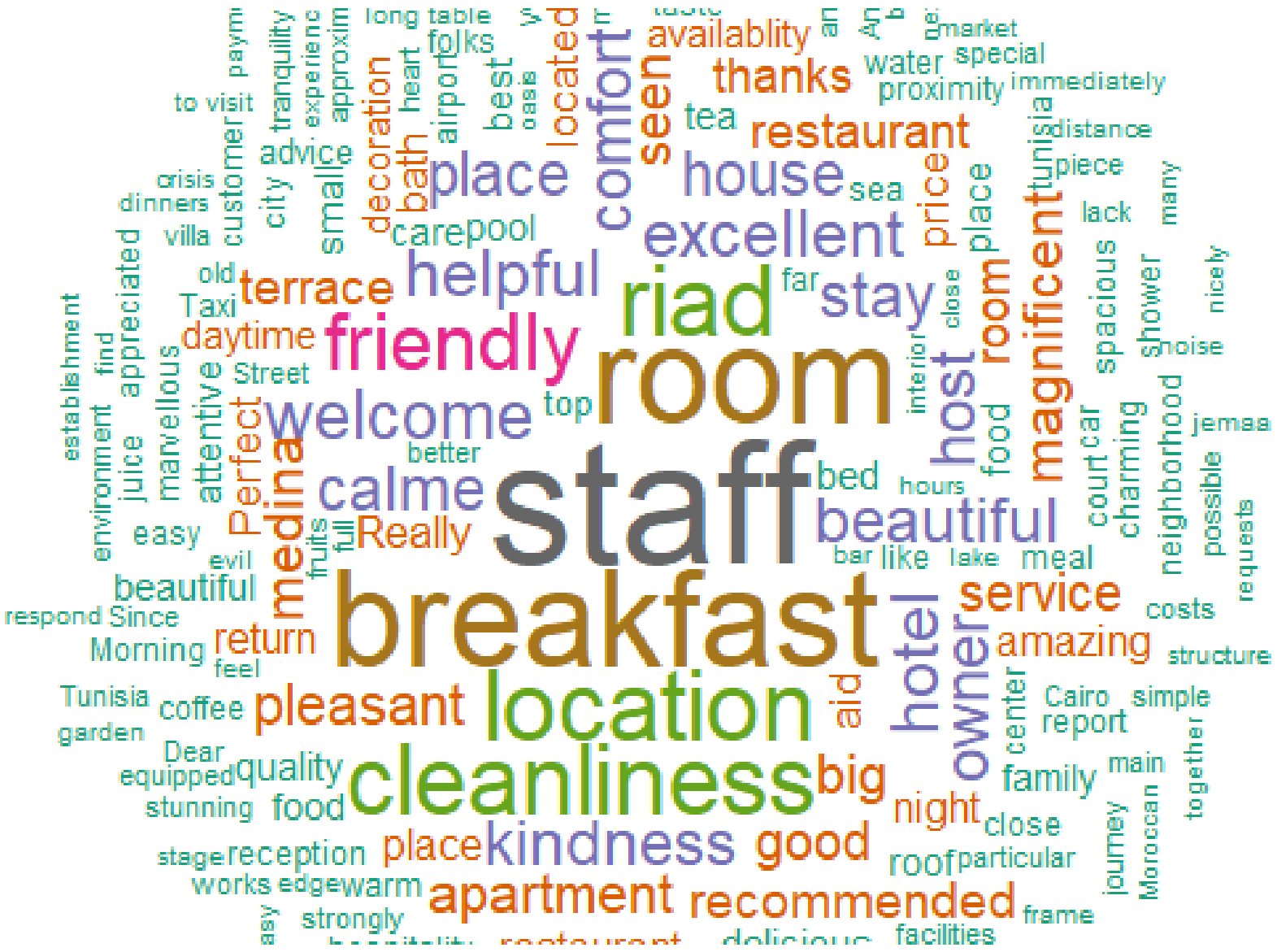

*Source: Authors from data extracted from the Web*

The word “staff”, the most frequent with 177 occurrences, is often associated with words with a positive connotation (helpful and sympathetic with correlation coefficients of 35%). The most frequent words besides staff are breakfast (129), associated with words with positive connotations (excellent and delicious with correlation coefficients of 25%) and location (92) associated, also, with positive feelings (central and medina with correlation coefficients of 21%). The sentiment analysis of the comments reveals the preponderance of words with positive connotations up to 92% for the three flagship destinations of North Africa (Graph 13). This large share of positive reviews may be due to the social proximity and empathy developed by tourists during their stays, hampering their willingness to provide negative reviews online to express their bad experiences (Pera et al., 2019; Tassiello et al., 2018).

**Graph 13: Word cloud of comments, according to their sentimental polarities, of the territorial tourist offer visible on the web of the flagship destinations of North Africa in 2020**

**Marrakech-Safi** **Tunis** **Cairo**

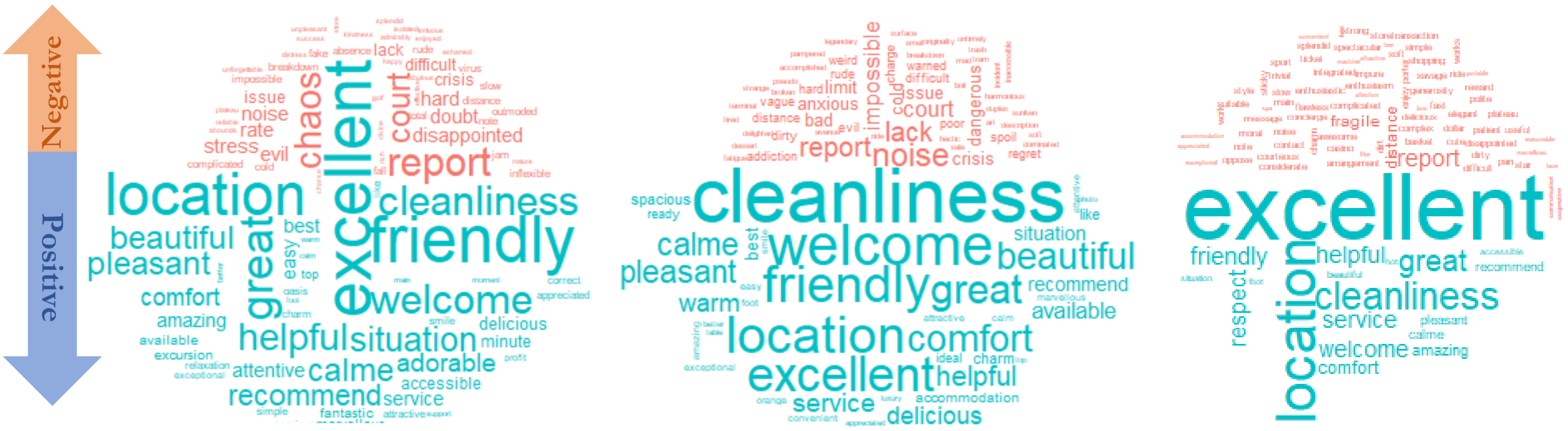

*Source: Authors from data extracted from the Web*

However, a differentiation exists in terms of appreciation within these regions. Indeed, French tourists have a more negative assessment relative to other nationalities and this, for the three territories. The same goes for the United Kingdom for the Moroccan region of Marrakech-Safi and the Tunisian Governorate of Tunis (respectively 91% and 89%) against the Germans who appreciate these destinations more (respectively 94% and 98%). Nationals also seem to appreciate North African destinations better than non-residents, particularly in the Moroccan region of Marrakech-Safi (Graph 14). Therefore, territorial tourism promotion should be differentiated according to the nationality of tourists who seem to have a differentiated appreciation and expectations. Indeed, extreme opinions seem to have a greater impact on overall satisfaction compared to moderate opinions (Nicolau et al., 2022).

**Graph 14: Ratio of positive words in comments on the territorial tourism offer visible on the web of North Africa's flagship destinations relative to the total of positive and negative words in 2020**

**Marrakech-Safi** **Tunis** **Cairo**

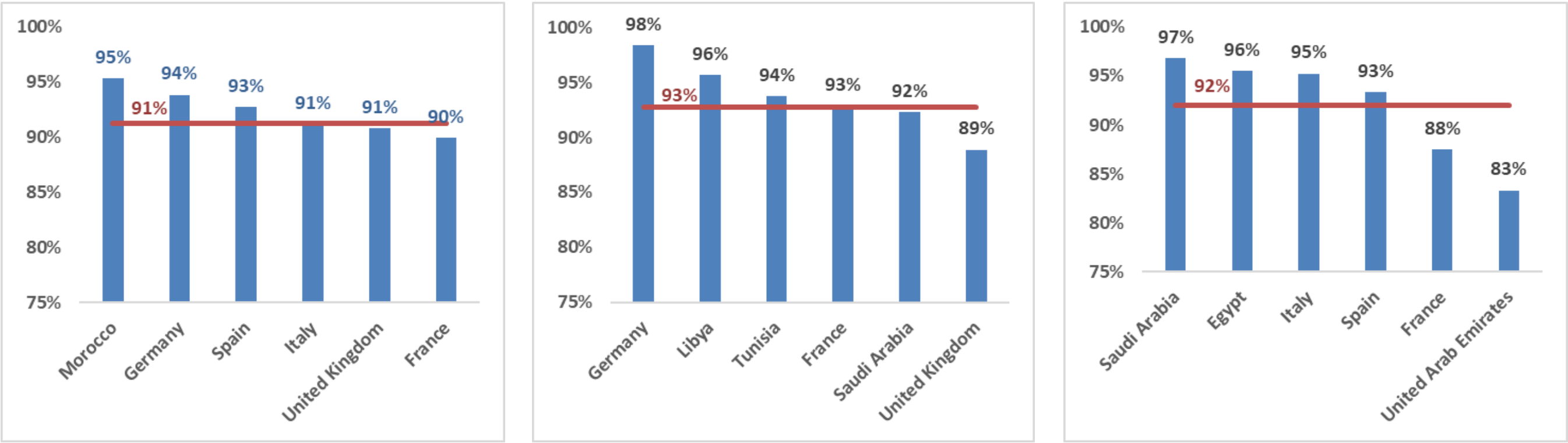

*Source: Authors from data extracted from the Web*

In this sense, this promotion should highlight the comparative advantages of the territories generating positive feelings as revealed in the comments. These include, in particular, the location in or near the Medina for Tunis, the Jamaa El Fna Square for Marrakech-Safi and the Museum and the Nile for Cairo. This concentration on the improvement of the failing attributes of the destinations makes it possible not to deteriorate the appreciation of other characteristics. Indeed, tourists tend to be more severe in their negative assessment of a specific characteristic, going so far as to undervalue other characteristics of the hotel (Mellinas et al., 2019). The most influential attributes concern "staff", "value for money" and "services" which can have asymmetric effects on the appreciation of the other attributes, in particular, "location" modifying the positive effect of halo (Lee et al., 2019) to a crown of thorns (Nicolau et al., 2020).

## 4. Conclusion

Thus, we have highlighted the opportunity of recent and detailed unconventional data from the tourism sector collected from the online accommodation platform « Booking.com ». These data allowed us to make a finer and more up-to-date analysis than that established by conventional data, particularly at the territorial level. Indeed, our analysis covered 68 territories of the six North African countries based on about forty variables covering 1852 accommodations. We analyzed the appreciation of the territorial tourist offer on the feedback of nearly 606000 tourists for all the territories of the six North African countries. We analyzed the sentiment of their comments by focusing on a sample of the most recent assessments of 10% of the accommodations of the most dynamic territories, namely the region of Marrakech-Safi, the governorate of Tunis and the mohafazat of Cairo.

It turns out that the accommodation offer of these destinations is very diversified covering types of accommodation (apartments, lodges, guest houses, …) which improve the authentic tourist offer of the territories[7]. Indeed, the unclassified offers from the territories of North Africa are better appreciated than the classified ones. Consequently, there is a need to support small accommodation structures in terms of funding and training given their importance in the attractiveness of territories. Also, any public policy in favor of tourism at the central or decentralized level should take into account these tourist actors who are not organized as a corporation, which puts them out of the negotiations, measures and tools put in place by public authorities and local communities.

These small structures often lack competitive and organizational capacities relative to large operators to face the major challenges of the sector, in particular, in times of shortage (phenomenon of cancellation, low occupancy rate, seasonality, etc.). It would therefore be wise to support these establishments in the sense of improving the visibility of their offers and sustaining their activities, or even supplementing it with services other than accommodation by drawing on the revealed expectations of the clientele. This is, in particular, the upgrading of equipment, in particular, in connectivity as noted through the expectations of tourists. National or regional support funds could be used to support professionals in this perspective with an offer adapted to small accommodation structures and support in putting together their application file. Indeed, small hosting companies have difficulties in obtaining support from government institutions to develop and grow (Mxunyelwa & Henama, 2019).

North African territories could also promote their offers marked by authenticity and sustainability to compete with European mass tourism destinations. Indeed, tourists are less demanding with regard to certain aspects of accommodation in favor of more ecological responsibility of hotel companies, in particular, in terms of water and energy management. They would be willing to pay conventional hotel prices for a green hotel while tolerating minor inconveniences (Y. Kim & Han, 2010). This is even more true for the new generations. Indeed, injunctive social norms have a significant indirect effect, via personal norms, on Gen Z's intention to choose an eco-friendly hotel (D'Arco et al., 2023). Also, the green labeling of certain small hotel structures could relatively counter certain shortcomings, particularly in terms of quality/price ratio, equipment and connectivity. This labeling is all the more important since green advertising can strengthen the confidence of European tourists (Nekmahmud et al., 2022) who are the main customers of North Africa.

[7] *The relationship between company size and customer experience is relatively complicated. An increase in the size of the company would improve the functional and emotional experience of the customer but would deteriorate the authentic experience (Ye et al., 2019).*

Web platforms could play an important role in this sense to support the promotion of territories, by using their referencing tools, and for the collection of tourist taxes for the benefit of local communities. Thus, the Regional Tourism Committee of New Aquitaine signed an agreement with Airbnb in March 2019 for the development of tourist activity in departments with strong untapped potential (Airbnb, 2019). That said, the web price index[8] would be a potential indicator for monitoring the economic activity of national destinations. Indeed, the variability of prices according to the booking date would suggest a link, among other things[9], with the attractiveness of the destination.

Thus, this work highlights the opportunities offered by "unconventional" data in the analysis and monitoring-evaluation of the tourism sector. Other possibilities are possible by exploiting other vectors of information (mobile telephony, geolocated route traffic, ...), in particular those of social media (Twitter, Youtube, Facebook, …) which have become important vectors of marketing and influence for the tourism sector (Gretzel, 2018). This calls for the creation of a multidisciplinary critical mass of experts at North African or even Mediterranean level within the framework of public (central and local)-private-university cooperation. This cooperation must be commensurate with the challenges and issues that arise, in particular, in terms of reducing the knowledge divide which penalizes the socio-economic emergence of developing countries and which accentuates the inequalities between its territories.

---

[8] *The extraction of prices according to the reservation date for several months would allow the construction of such index. Such work could be envisaged in the medium to long term, the time to constitute an exploitable series.*
[9] *This development could also be due to the early sale of the cheapest offers during the first two months or the late scheduling of a major event.*